\documentclass[journal]{IEEEtran}
\usepackage{mathtools}
\usepackage{amsmath}
\ifCLASSINFOpdf
\else
\fi
\begin{document}
%
\title{Discriminative Dimensionality Reduction using Deep Neural Networks for Clustering of LIGO data}
%
%
%

\author{Sara Bahaadini, Yunan Wu, Scott Coughlin, Michael Zevin, and Aggelos K. Katsaggelos 
\thanks{S. Bahaadini, Y. Wu and A. K. Katsaggelos are with the Image and Video Processing Laboratory (IVPL), Department of Electrical and Computer
Engineering, Northwestern University, Evanston, IL, 60208, USA. e-mail: sara.bahaadini@gmail.com; yunanwu2020@u.northwestern.edu; a-katsaggelos@northwestern.edu}
\thanks{S. Coughlin is with the Center for Interdisciplinary Exploration and Research in (CIERA), Northwestern University, Evanston, IL, 60208, USA. e-mail: scottcoughlin2014@u.northwestern.edu}
\thanks{M. Zevin is with the Kavli Institute for Cosmological Physics, Enrico Fermi Institute, University of Chicago, Chicago, IL, 60637, USA.
e-mail: michael.j.zevin@gmail.com}}

%
%

\markboth{Journal of The IEEE Transactions on Neural Networks and Learning Systems}%
{Shell \MakeLowercase{\textit{et al.}}: Discriminative Dimensionality Reduction using Deep Neural Networks for Clustering of LIGO data}
%



\maketitle

\begin{abstract}
In this paper, leveraging the capabilities of neural networks for modeling the \textit{non-linearities} that exist in the data, we propose several models that can project data into a \textit{low dimensional}, \textit{discriminative}, and \textit{smooth} manifold. The proposed models can transfer knowledge from the domain of known classes to a new domain where the classes are unknown. A clustering algorithm is further applied in the new domain to find potentially new classes from the pool of unlabeled data. The research problem and data for this paper originated from the Gravity Spy project which is a side project of Advanced Laser Interferometer Gravitational-wave Observatory (LIGO). The LIGO project aims at detecting cosmic gravitational waves using huge detectors. However non-cosmic, non-Gaussian disturbances known as ``glitches'', show up in gravitational-wave data of LIGO. This is undesirable as it creates problems for the gravitational wave detection process. Gravity Spy aids in glitch identification with the purpose of understanding their origin. Since new types of glitches appear over time, one of the objective of Gravity Spy is to create new glitch classes. Towards this task, we offer a methodology in this paper to accomplish this. 
\end{abstract}


%
\IEEEpeerreviewmaketitle

\section{Introduction}
Over recent years, deep learning has shown impressive results in various machine learning areas ~\cite{lecun2015deep}. However, deep neural networks require a large amount of labeled training data to work effectively, and such data are not easily acquired for many problems due to its high acquisition and labeling cost and unavailability. Semi-supervised learning addresses this problem by exploiting unlabeled data when a small subset of labeled data is available \cite{zhu_introduction_2009}. In many applications, there are plenty of unlabeled data available which are usually inexpensive to obtain and can alleviate the scarcity of labeled data \cite{chapelle2006semi}. 

One field where semi-supervised learning has proven  exceptionally useful has been in the classification of transient noise in gravitational-wave detectors \cite{cuoco_enhancing_2020,marianer_semisupervised_2021}. 
Gravitational waves are the minuscule perturbations in the fabric of spacetime predicted as a byproduct of Einstein's theory of general relativity over a century ago. 
Starting with the first observation of gravitational waves in 2015~\cite{2016PhRvL.116f1102A}, which were produced by the merger of two black holes, the LIGO~\cite{2015CQGra..32g4001L} and Virgo~\cite{2015CQGra..32b4001A} interferometer network has observed dozens of other gravitational-wave events originating from the merger of black holes and neutron stars. 

To observe such astrophysical events, the LIGO and Virgo interferometers need to be sensitive to changes of distance of $\mathcal{O}(10^{-19})$ meters. 
Thus, despite sophisticated instruments and techniques designed to isolate the interferometer system, the detectors are susceptible to a variety of instrumental and environmental sources of noise. 
These transient noise sources, colloquially referred to as ``glitches'', come in many time-frequency morphologies, occur at a rate much higher than the rate of detectable astrophysical events, and can masquerade as true astrophysical events. 
Their classification and characterization, as well as their subsequent removal from the instrument data entirely, are a paramount task of gravitational-wave scientists.  

In \cite{zevin2017gravity}, we presented the Gravity Spy project\footnote{www.gravityspy.org}, which combines human classification schemes with machine learning to characterize noise sources within the LIGO datastream. 
Gravity Spy relies on deep learning techniques \cite{saraicassp}, using a convolutional neural network algorithm trained on a pre-labeled training set of 22 classes of glitches. 
This data is then exported to human volunteers, who perform their own classification task on the machine-labeled data. 
As of June 2021, the project has received over 5 million classifications from over 25000 registered volunteers. 
We have made the training data set publicly available \cite{sarajournal1}, and created an interactive database for LIGO scientists to query and further analyze the classification results for all the unlabeled data in the Gravity Spy testing set. 

One of the unique and challenging qualities of this particular classification task is the variation of the dataset domain over time. 
Gravitational-wave detectors are continuously-evolving instruments, and are periodically upgraded to increase their sensitivity. 
Therefore, over time certain classes of glitches disappear from the data whereas new morphological classes arise. 
As part of the Gravity Spy project, a semi-supervised clustering algorithm was developed~\cite{direct}, which was used to morphologically classify unlabeled glitches and identify new clusters representing morphologically distinct classes. 
This tool was also made generally accessible so that both volunteers of the Gravity Spy project and gravitational-wave scientists could query and build sets of glitches with similar morphological characteristics. 

\begin{figure*}[ht]
\centering
\includegraphics[width=0.8\textwidth]{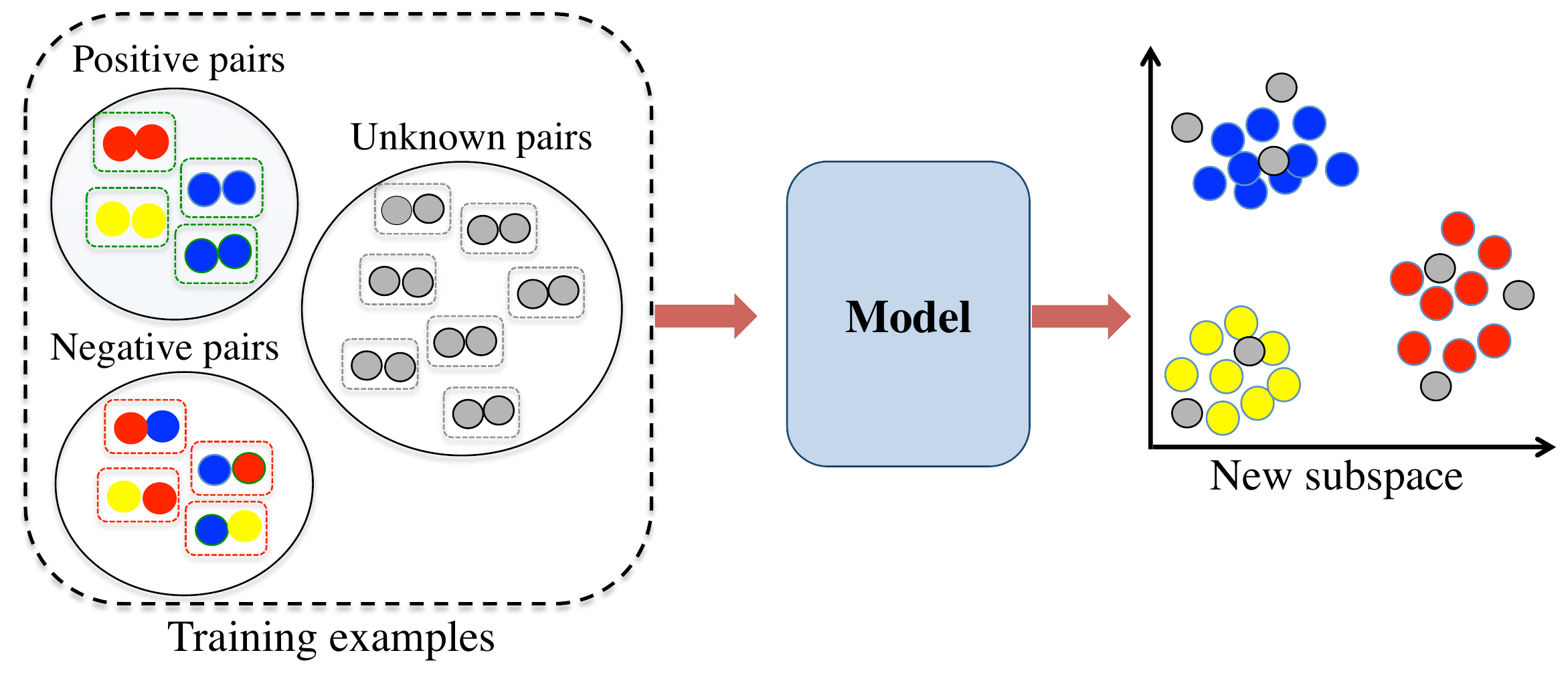}
\caption{Using both labeled and unlabeled pairs, the model learns  nonlinear functions that map the original samples into a new subspace in which samples of a class tend to be close to each other while they are mapped far from the samples of the other classes.}
\label{fig:arch}
\end{figure*}

In this paper, we are presenting a framework for dimensionality reduction that maps samples from a high dimensional feature space to a discriminative low dimensional space using deep neural networks. Our models have two main parts, a supervised  or discriminative part that needs labeled data and an unsupervised part which regularizes the supervised part and boost the overall performance of the model. First we develop an auto-encoder based model \cite{meng_relational_2017} that integrates into the model we suggested in~\cite{direct}. This enables the model to learn from both labeled and unlabeled data in a joint manner. 

We also propose a model that uses a technique called ``Virtual Adversarial Training" (VAT)~\cite{VAT}. The main idea in VAT is derived from adversarial training where the goal is to find the adversarial direction or perturbation so as to protect a machine learning model from being fooled by adversarial examples. In the machine learning literature \cite{goodfellow_generative_2014}, adversarial examples are referred to some special types of input of machine learning models (including neural networks but not limited to) that result in an incorrect output from the model. Adversarial training is one of the approaches that try to add many of such adversarial examples to the training set with their correct labels and explicitly teach the model how not to be fooled. Here, we are inspired by the regularization effect of VAT to further improve the discriminative dimensionality reduction model. We evaluate the performance of the new models with a clustering task. The scenario that we are focusing on is similar to transfer learning. The classes that are used for training the neural network are totally disjoint from the classes we perform clustering on.

Our experimental results show that the proposed model uses unlabeled data can perform better than the fully supervised model such as~\cite{direct}. Between the auto-encoder based and VAT based dimensionality reduction algorithms, the VAT based one performs better, which shows the effectiveness of regularizing the neural network by adding the adversarial perturbation to the training data.

The rest of the paper is organized as follows. In Section 2, we explain the proposed model architectures, including the auto-encoder based model and the virtual adversarial based model. In Section 3, we present data collections and compare experimental results among different models. The final Section is the conclusion.

\section{Dimensionality Reduction Models using Neural Networks}

\subsection{Definitions}
The training of the models is performed with pairs of images in the same way as in \cite{direct}. More specifically, training data consist of positive, negative, and unknown pairs based on whether the two samples of a pair belong to the same, different, or an unlabeled set, respectively (see Figure~\ref{fig:arch}). The positive and negative pairs are used for training the discriminative component in a supervised manner. The unsupervised component is trained in an unsupervised fashion by all the pairs including positive and negative pairs. The nonlinear function, learned through training the model, project the samples into a new discriminative subspace where samples with positive relations are mapped close to each other and samples with negative relations far apart.

We are given a set of training pairs $X = \big\{ (x_1^i,x_2^i)| 1 \le i \le N  \big \}$, with $x_j^i \in \mathbf{R}^m$, and a set $Y = \{y^i| 1 \le i \le N , y \in \{ p,n,u \}\}$ which contains the labels that specify the relation between the samples of each pair. A relation can be positive $p$, negative $n$ or unlabeled $u$. Positive relation indicates that the two samples belong to the same class and negative relation indicates the opposite. The relations for the unlabeled pairs are unknown. We assume that $L$ of these pairs are labeled and $U$ are unlabeled, so that $N=L+U$.

\begin{figure*}[bt]
\centering
\includegraphics[width=.95\textwidth]{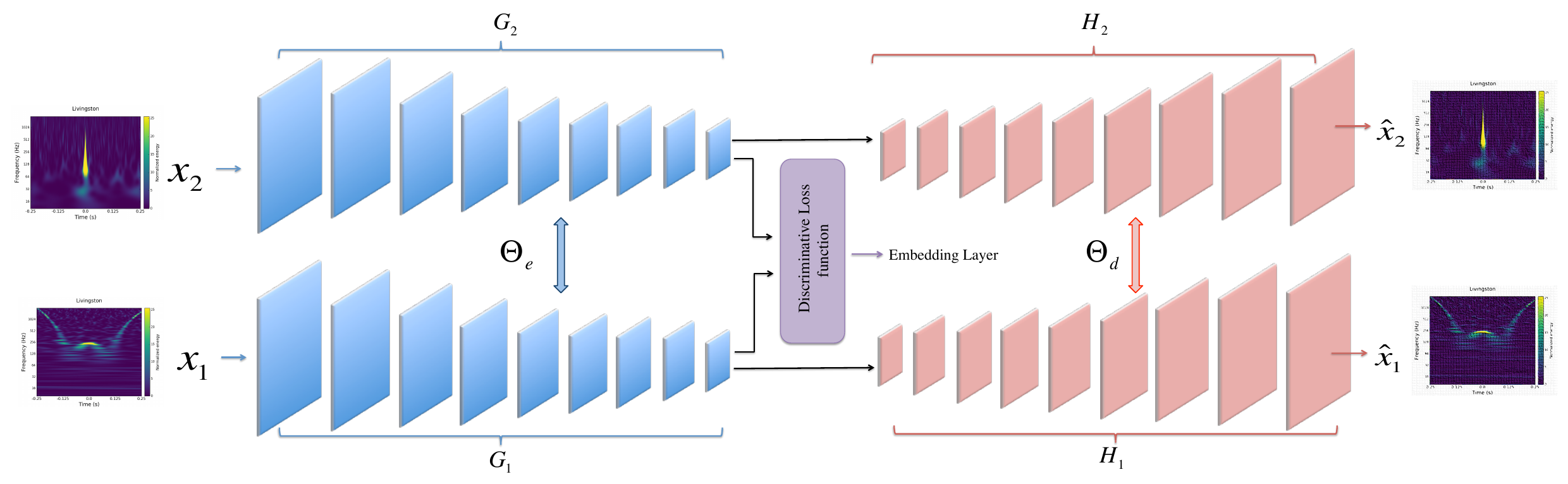}
  \caption{The left subnetworks $G_1$ and $G_2$ perform supervised learning for the model and the right subnetworks $H_1$ and $H_2$ help unsupervised learning. The contrastive layer makes the supervised part of the model to learn a discriminative subspace.}
  \label{fig:sara-seven}
\end{figure*}

\subsection{Auto-encoder based Model}
\label{sec:seven}
The overall architecture of the proposed model is illustrated in Figure \ref{fig:sara-seven}. The input pair is given to two neural networks denoted by $G_1$ and $G_2$ with shared weights and parameters $\Theta_e$. They form the supervised part of the model and project the input samples to a new discriminative space where samples with positive relations are close to each other and samples with negative relations are far from each other. As the weights of $G_1$ and $G_2$ are shared, both subnetworks define the same nonlinear mapping function, denoted by $g(.;\Theta_e)$. To make the new representation have such discriminative property, a layer is added at the top of the networks $G_1$ and $G_2$ that estimates the distance between the two input samples in the new space denoted by $d(x_1, x_2;\Theta_e)$. Function $d$ can be an arbitrary distance metric such as Euclidean or cosine distance in the new subspace. This function will be considered as a metric distance function which networks $G_1$ and $G_2$ are supposed to learn.

The outputs of $G_1$ and $G_2$, i.e., new representations, are inputs to two other subnetworks denoted by $H_1$ and $H_2$ with shared weights $\Theta_d$. They are from the unsupervised part of the model and map back the samples from the new space to the original space. In other words, they work as decoders for the encoders $G_1$ and $G_2$ and their corresponding mapping function $h(.;\Theta_d)$ reconstructs the input pairs. The outputs of $H_1$ and $H_2$ (shown as $\hat{x}_1$ and $\hat{x}_2$ in Figure \ref{fig:sara-seven}) are the reconstruction of the corresponding inputs $x_1$ and $x_2$. 

The unsupervised subnetworks are considered to be auxiliary networks for the supervised subnetworks of the model. These auxiliary subnetworks act as regularizers for the supervised part. They enable the supervised learning process to benefit from the unlabeled data by imposing some constraints on the feature-learning process required for the specific target task. 

The new subspace obtained from the encoders $G_1$ and $G_2$ should be discriminative so that samples from different classes be  easily separable. 
However, the discriminative property is not enough for semi-supervised settings where the relation of some pairs is not available. To exploit the information of such samples, we impose another constraint. The desired representation should maintain the information in the inputs as much as possible. This can be obtained by imposing a reconstruction constraint. We combine both constraints into a unified loss function $\mathcal{J}(X,Y;{\Theta _e},{\Theta _d})$ which includes 
$\mathcal{J_D}(X,Y;{\Theta _e})$, the supervised loss for labeled data, and $\mathcal{J_G}(X;{\Theta _e},{\Theta _d})$, the unsupervised loss for all data. $\mathcal{J_D}(X,Y;{\Theta _e})$ can be estimated for the $L$ labeled pairs as follows:
\begin{equation}
\mathcal{J_D}(X,Y;{\Theta _e})=\sum_{1\le i\le L} j_d(x_1^i,x_2^i,y^i;\Theta_e),
\label{eq:eq3}
\end{equation}
where $j_d(x_1^i,x_2^i,y^i;\Theta_e)$ indicates the discriminative loss for the pair $(x_1^i,x_2^i)$ in the new subspace, and $y^i$ is the corresponding label for the pair. It can be defined with a constrastive loss function as follows:

\begin{align}
\begin{split}\label{eq:ll}
j_d(x_1^i,x_2^i, y^i;\Theta_e) = y^id({x_1^i},{x_2^i};{\Theta _e}) \\
+ (1-y^i)[1-d({x_1^i},{x_2^i};{\Theta _e})].
\end{split}
\end{align}

Contrastive loss function and its variants are commonly used in siamese networks \cite{qi2016sketch}. It penalizes the distance between positive samples and also the similarity between negative ones. We define the distance metric function $d$ as follows based on the cosine similarity; however it can be an arbitrary distance function.

\begin{align}
\begin{split}\label{eq:}
d({x_1},{x_2};{\Theta _e})=1-cos(g(x_1;{\Theta _e}),g(x_2;{\Theta _e})),
\\ 0 \le d({x_1},{x_2};{\Theta _e}) \le 1.
\end{split}
\end{align}

Similarity function $cos(a,b)$ measures the cosine similarity between its two vectors $a$ and $b$ as:
\begin{equation}
cos(a,b) = \frac{a^Tb}{\left \| a \right \|\left \| b \right \|}.
\end{equation}

The generative loss $\mathcal{J_G}(X;{\Theta _e},{\Theta _d})$ is estimated by all of $L+U$ pairs according to  
\begin{equation}
\mathcal{J_G}(X;{\Theta _e},{\Theta _d}) = \sum_{1\le i\le L+U} j_g(x_1^i,x_2^i;\Theta_e,\Theta_d),
\label{eq:}
\end{equation}
\noindent
where $j_g(x_1,x_2)$ indicates the generative loss of the pair $(x_1,x_2)$ based on their reconstructions from the their hidden representations,i.e.,

\begin{align}
\begin{split}\label{eq:}
j_g(x_1^i,x_2^i;\Theta_e,\Theta_d) = 
\|h(g(x_1^i;\Theta_e))-x_1^i\|_2 
 + \\ \|h(g(x_2^i;\Theta_e))-x_2^i\|_2. 
\end{split}
\end{align}

By combining the generative and discriminative losses, the optimization problem of the whole model is written as:
\begin{multline}
 \mathop{} \limits_{{\Theta _d},{\Theta _e}} \mathcal{J}(X,Y;{\Theta _e},{\Theta _d}) =  \sum\limits_{i = 1}^L {{j_d}(x_1^i,x_2^i,y^i;{\Theta _e})}  + \\
 \alpha \sum\limits_{i = 1}^{L + U} {{j_g}(x_1^i,x_2^i;{\Theta _e},{\Theta _d})}  +   \\
 \beta (\left\| {{\Theta _e}} \right\|_2 + \left\| {{\Theta _d}} \right\|_2),
 \label{eq:eqj}
\end{multline}

where $\|\Theta_e\|$ and $\|\Theta_d\|$ are the regularization terms on the parameters of the networks $G_i$ and $H_i$ controlled by the regularization parameter $\beta$. Parameter $\alpha$ controls the trade off between the discriminative and generative objectives.
In the next step, $U$ are passed through the $G_1$ subnetwork, in other words $z^k = g(x^k;{\Theta _e})$, where $1\le k \le U$. Then a clustering method such as k-means will be applied on the transformed data ${\Big\{z^k}\Big\}_{k=1}^{U}$.

\begin{figure}[ht]
    \centering
    \includegraphics[width=0.4\textwidth]{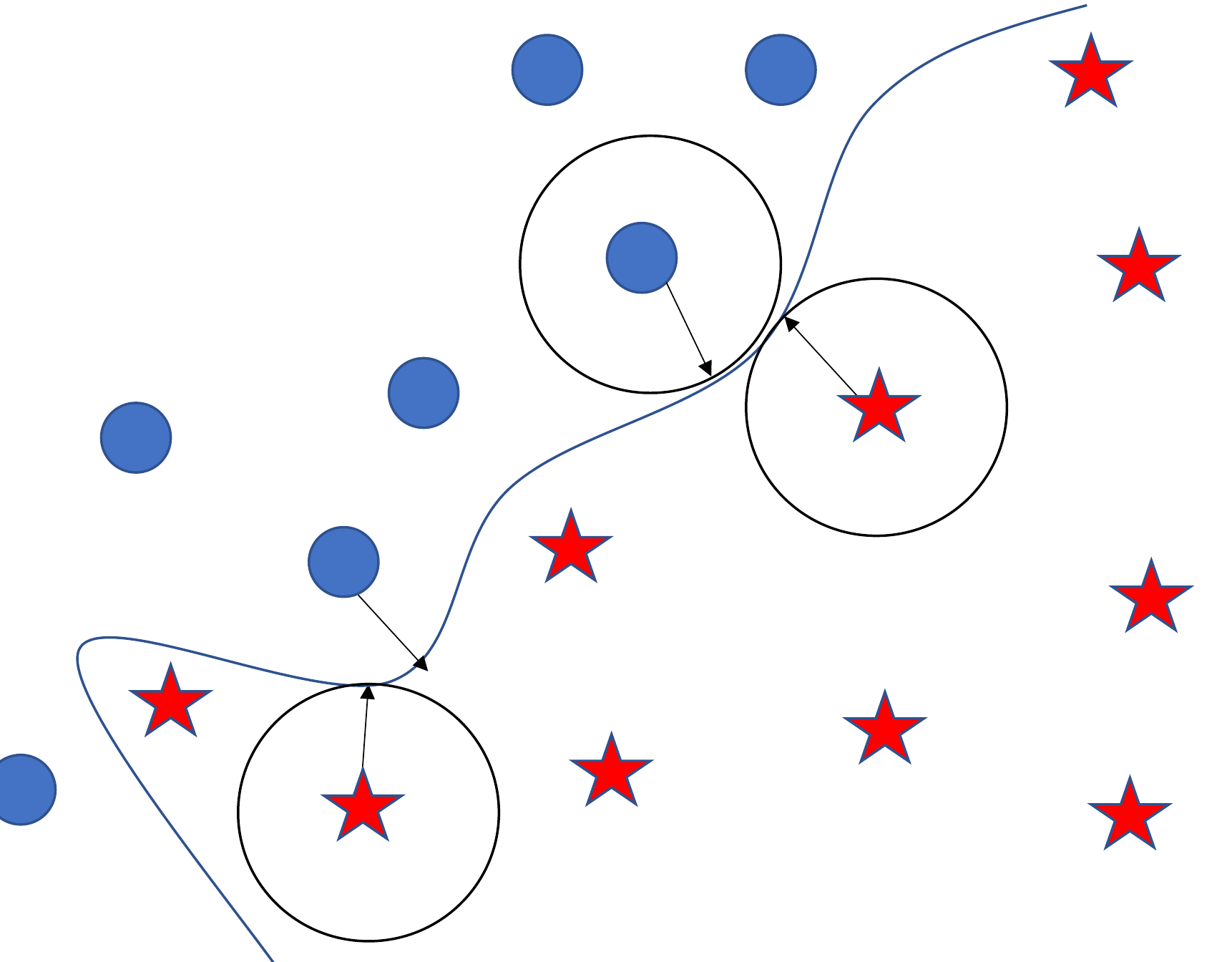}
    \caption{The regularization effect of increasing the local distributional smoothness.}
    \label{fig:vatshape}
\end{figure}

\subsection{Virtual Adversarial based Model}
We do not want a large shift in the model's output by perturbing the model's input a little. Making the model robust to such perturbation can be considered as a means of regularizing the model.
Such smoothing and regularization should increase the generalization of the model as the model tries to predict similar outputs for unseen samples close to the ones present in the training set (see Figure~\ref{fig:vatshape}). 

\begin{figure*}[ht]
\centering
\includegraphics[width=.95\textwidth]{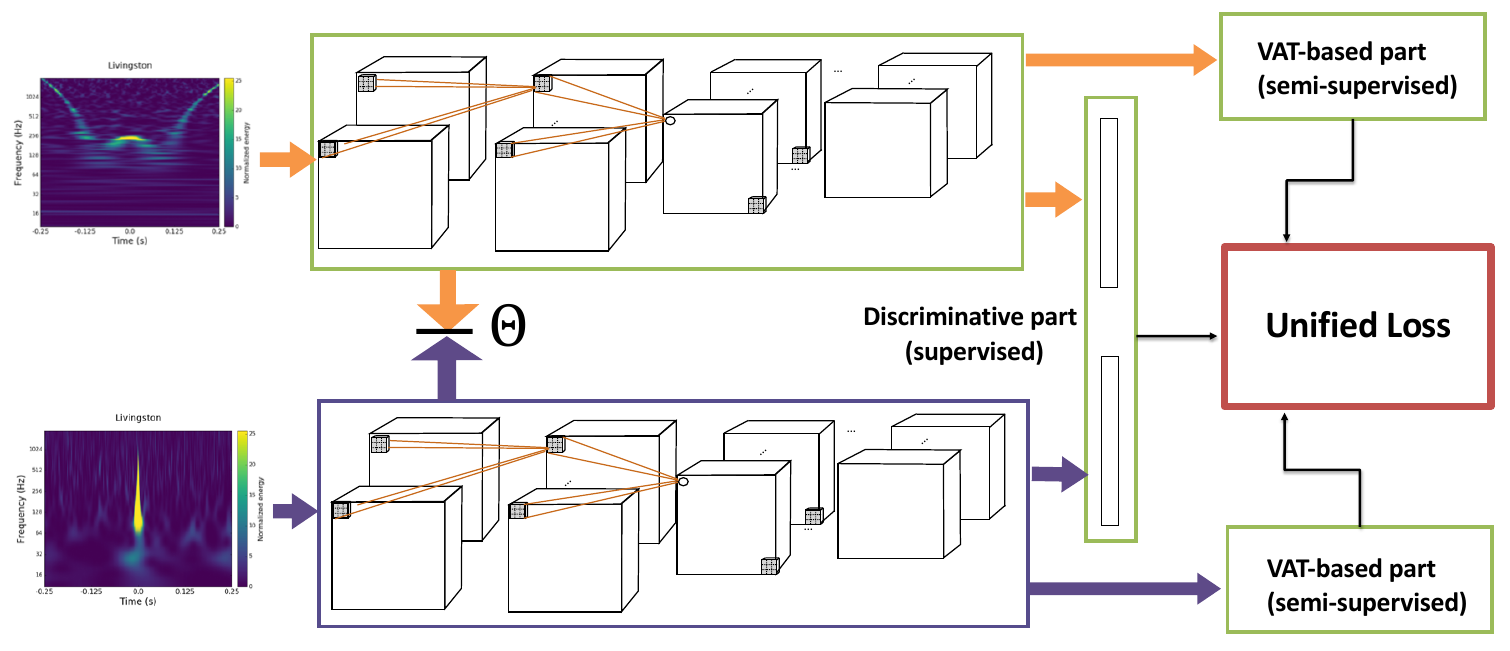}
  \caption{The two parallel neural networks share their parameters and project the samples to an abstract level through convolutional, max pooling and fully connected layers. The supervised part of the loss function encourages discriminativity and the VAT-based loss function increase. }
  \label{fig:vat}
\end{figure*}

Similarly to auto-encoder based modeling, we have two parallel neural networks that share their parameters in a Siamese-like architecture. They form an encoder and project samples to a new low dimensional feature space that we want to be ``discriminative" and ``smooth" (see Figure~\ref{fig:vat}). We heretofore refer this model as the VAT-based plus discriminative model. 

This can be obtained by imposing a reconstruction constraint. We combine both constraints into a unified loss function as:
\begin{equation}
\mathcal{J}(X,Y;{\Theta}) = \alpha\mathcal{J_D}(X,Y;{\Theta}) + (1-\alpha)\mathcal{J_V}(X;{\Theta}), 
\label{eq:eq4}
\end{equation}
where $\mathcal{J_D}(X,Y;{\Theta})$ indicates the supervised loss for labeled data, and $\mathcal{J_V}(X;{\Theta})$ the VAT semi-unsupervised loss.
The latter is estimated by all $L+U$ pairs according to 
\begin{equation}
\mathcal{J_V}(X;{\Theta}) = \sum_{1\le i\le L+U} l_v(x_1^i,x_2^i;\Theta),
\label{eq:vat_loss}
\end{equation}
\noindent
where $l_v(x_1^i,x_2^i;\Theta)$ indicates the VAT based loss of the pair $(x_1^i,x_2^i)$. It is defined as
\begin{equation}
l_v(x_1^i,x_2^i;\Theta) = 
l_{v1}(x_1^i;\Theta) + l_{v2}(x_2^i;\Theta), 
\label{eq:vat_pair_loss}
 \end{equation}
\noindent
where 
\begin{align*}
{l_{v1}}(x_j^i;\Theta) = KL(f(x_1^i;\Theta),f(x_1^i + {r_{vr-adv}^i};\Theta)),    
\end{align*}
\begin{equation}
{r_{vr-adv}^{i}} = \mathop {\arg \max KL(}\limits_{r;{{\left\| r \right\|}_2} < \varepsilon } f(x_1^i;{\Theta}),f(x_1^i + r;{\Theta})),
\label{eq:11}
\end{equation}

\noindent
where $KL(.,.)$ is the Kullback–Leibler divergence between the two output distributions, and $\varepsilon$ is a tiny positive number. $r_{vr-adv}$ is the virtual adversarial vector that defines the direction in the input vector that causes the maximum change in the output distribution. Eq.(\ref{eq:11}) also applies to $l_{v2}$. $r_{vr-adv}$ does not have a closed from solution and is therefore approximated by:
\begin{equation}
{r_{vr-adv}} = \varepsilon \frac{\nabla_r KL(f(x_1^i;{\Theta}),f(x_1^i + r;{\Theta}))}{{\left\| \nabla_r KL(f(x_1^i;{\Theta}),f(x_1^i + r;{\Theta})) \right\|}},
\end{equation}
\noindent
where
\begin{equation}
r \sim N(0,\frac{\varepsilon }{{\sqrt {D_x} }}I),    
\end{equation}
\noindent
that is a normally distributed random vector added to the input of the neural network as the target perturbation. It has the same dimension as the input dimension $D_x$, and $I$ is a $D_x \times D_x$ identity matrix.
In practice this gradient vector is estimated using back-propagation on the computation graph of the neural network. More details on this approximation can be found in \cite{VAT}.

\section{Experiments}
\label{sec:vat_exp}
\subsection{Dataset}

\begin{table*}[ht]
\centering
\caption{Breakdown of morphological categories in the Gravity Spy training set, indicating the number of training set samples of each class originating from the Livingston and Hanford detectors. Note that some of the glitches are detector dependent.}
\label{tbl:class_numbers}
\begin{tabular}{|l|r|r|l|r|r|}
\hline
\textbf{Class              } &  Hanford &  Livingston & \textbf{Class              } &  Hanford &  Livingston \\
 \hline\hline
\textbf{1080Lines          } &      327 &           0 & \textbf{No Glitch        } &       65 &          52 \\ 
\textbf{1400Ripples        } &        0 &          81 & \textbf{None of the Above} &       51 &          30 \\
\textbf{Air Compressor     } &       55 &           3 & \textbf{Paired Doves     } &       27 &           0 \\
\textbf{Blip               } &     1452 &         369 & \textbf{Power Line       } &      273 &         176 \\
\textbf{Chirp              } &       28 &          32 & \textbf{Repeating Blips  } &      230 &          33 \\
\textbf{Extremely Loud     } &      266 &         181 & \textbf{Scattered Light  } &      385 &          58 \\
\textbf{Helix              } &        3 &         276 & \textbf{Scratchy         } &       90 &         247 \\
\textbf{Koi Fish           } &      517 &         189 & \textbf{Tomte            } &       61 &          42 \\
\textbf{Light Modulation   } &      511 &           1 & \textbf{Violin Mode      } &      141 &         271 \\
\textbf{Low Frequency Burst} &      166 &         455 & \textbf{Wandering Line   } &       42 &           0 \\
\textbf{Low Frequency Lines} &       79 &         368 & \textbf{Whistle          } &        2 &         297 \\
\hline
\end{tabular}

\end{table*}

\begin{figure*}[ht]
\centering
\includegraphics[width=18cm,height=16cm]{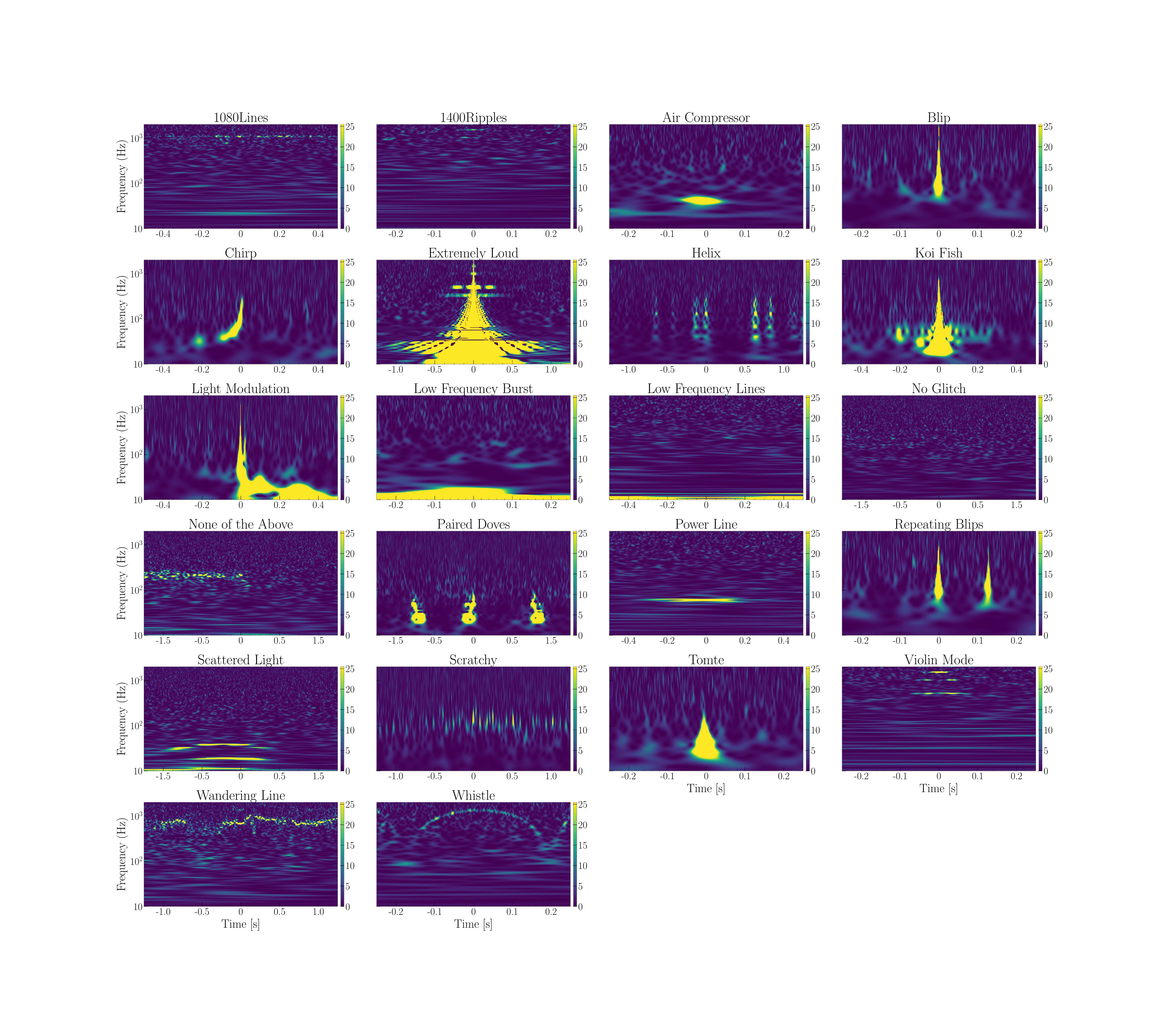}
\caption{Omega Scan images for example members of each class within the Gravity Spy dataset. From top left to bottom right; row one: 1080Lines, 1400Ripples, Air Compressor, Blip; row two: Chirp, Extremely Loud, Helix, Koi Fish; row three: Light Modulation, Low Frequency Burst, Low Frequency Lines, No Glitch; row four: None of the Above, Paired Doves, Power Line, Repeating Blips; row five: Scattered Light, Scratchy, Tomte, Violin Mode; row six: Wandering Line, Whistle. Chirp is not strictly a glitch but it is an important category as real gravitational waves can appear in our data stream and the example of None of the Above is only one example of the various forms belong to this "class".}
\label{fig:all_glitches}
\end{figure*}

We use Version 1.2 of the Gravity Spy dataset, which improves upon Version 1.0 used in~\cite{sarajournal1} (available at~\cite{bahaadini_sara_2018_1476156}). 
The improvement comes from removing training set samples that were either incorrect, or poor representations of their respective morphological class. 
The training set now consists of a total of $7932$ glitches from both the LIGO-Livingston and LIGO-Hanford detectors. 
These glitches are grouped into 22 classes, with exact proportions shown in Table~\ref{tbl:class_numbers}. 
As can be seen from the table, the classes do not contain equal numbers of training samples, since not every class occurs in both detectors. 
Spectrograms of each class are shown in Figure~\ref{fig:all_glitches}. For more information on the names and potential causes of each glitch, see Section 2 of~\cite{sarajournal1}.

We use normalized mutual information (NMI) as the evaluation metric for the clustering task. It is defined as
\begin{equation}
NMI = \frac{I(Y; \hat{Y})}{\sqrt{H(Y) \times H(\hat{Y})}},
\end{equation}
where Y, $\hat{\rm Y}$, H, and I are the true clusters, the predicted clusters, the entropy and the mutual information, respectively. Notice that $0\leq $ NMI $\leq 1$, where the lower bound is achieved for $y$ and $\hat{y}$ independent, while the upper bound for $y = \hat{y}$.
\subsection{Experimental Settings}
To evaluate the effectiveness of the suggested dimensionality reduction technique, we perform clustering on the new feature space.
We use the first $10$ classes for training the deep neural network and the remaining $10$ classes for the clustering step. Therefore the classes used for training the model are disjoint from the classes we apply clustering on.

\subsection{The VAT based Model vs.The Auto-encoder based Model }
\label{sec:seven_exp}
First we compare the performance of the auto-encoder based model with the fully supervised one proposed in~\cite{direct}. Table~\ref{tab:res1} shows the performance of this model compared to the raw feature space, Principal component analysis (PCA) and fully supervised model (DIRECT).

As expected, the auto-encoder based model that has both unsupervised and discriminative components works the best. As the fully supervised or discriminative model uses the labeled information, it provides us with a more discriminative feature space that is shown to be more efficient for clustering than the raw features and the features obtained from PCA.

We have compared the performance of the VAT-based dimensionality reduction model with DIRECT~\cite{direct}, auto-encoder based dimensionality reduction models and the model using raw features. The results are presented in Table~\ref{tab:res1}. As we can see the VAT based model performs the best which confirms the effectiveness of the use of virtual adversarial training regularization effect.

\begin{figure}[h!]
    \centering
    \includegraphics[width=0.4\textwidth]{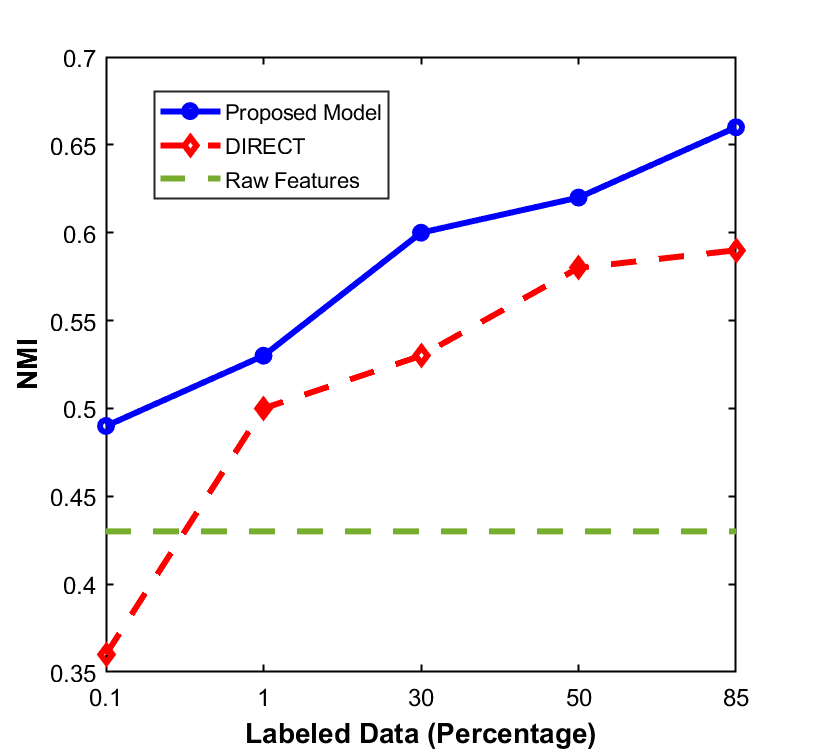}
    \caption{The NMI of the clustering problem for various features. The x-axis shows the percentage of labeled data used for training the VAT-based dimensionality reduction model. The model is further used as a feature extractor to project samples into a smooth, low dimension, discriminative feature space where the clustering algorithm can work more efficiently. }
    \label{fig:vat_exp2}
\end{figure}

In another set of experiments, we investigate the effect of using the various percentages of labeled data for training the VAT-based dimensionality reduction method. As can be seen in Figure~\ref{fig:vat_exp2}, as more labeled data is utilized by the network for training the obtained low dimensional feature space works better for clustering. Also, the VAT-based method compared to DIRECT~\cite{direct} works better which indicates that a smooth manifold is obtained by the VAT-based model which is better for clustering.
We also performed a series of pair classification experiments in a setting similar to the transfer learning one we discussed. Again, here, the set of classes used for training the model is disjoint from the set of classes the test pairs are configured from. The target task is to classify pairs to either positive or negative, belonging to same classes in the former case and different classes in the latter case. The results of this experiment are shown in Table~\ref{tab:vat_exp2}.

\begin{table}[h]
\centering
\caption{Accuracy of pair classification. The first row shows the percentage of labeled data used for training the neural network model. As expected the more the labeled data, the higher the accuracy.}
\label{tab:vat_exp2}
\begin{tabular}{c|c|c|c|c|c}
Labeled data (\%) & 0.1 & 1 & 30 & 50 & 85 \\ \hline
Accuracy (\%)& 49.07 & 49.07 & 79.12 & 79.49 & 82.62
\end{tabular}
\end{table}

\begin{table}[h!]
\centering
\caption{Comparison of the performance of the proposed model.}
\label{tab:res1}
\begin{tabular}{l|ll}
Methods       & NMI \\ \hline
Raw features         &  $0.43$     \\
PCA        &  $0.43$     \\
DIRECT~\cite{direct} &  $0.53$     \\
Auto-encoder based model plus discriminative       &$0.55$ \\
VAT based model plus discriminative & $\textbf{0.61}$
\end{tabular}
\end{table}

\section{Conclusion}
We presented a dimensionality reduction model that can use both labeled and unlabeled data to provide a discriminative and robust feature space. We used an auto-encoder structure and added a siamese architecture to encourage desired properties. Furthermore, we improved the first architecture by proposing a more sophisticated way of using unlabeled data to further improve the discriminative feature space by creating a smooth manifold. Using the regularization effect of virtual adversarial training (VAT), a smooth manifold which tends to assign to the neighborhood of a training sample the same label as the training sample is developed. The experiments were performed on two tasks: clustering and pair classification. Both tasks are performed in a transfer learning setting where the training and testing classes are disjoint. The results demonstrated that the VAT based dimensionality reduction outperforms all other models. This shows the effectiveness of projecting the samples into a low dimensional feature space that is discriminative (with contrastive loss) and smooth (with VAT based loss) using deep neural networks (non-linear modeling).


%



\section*{Acknowledgment}
 Gravity Spy is partly supported by the National Science Foundation, award INSPIRE 15-47880.

\ifCLASSOPTIONcaptionsoff
  \newpage
\fi



%



\bibliographystyle{IEEEtran}
\bibliography{paper}

 \newcommand{\noop}[1]{}
\begin{thebibliography}{10}
\providecommand{\url}[1]{#1}
\csname url@samestyle\endcsname
\providecommand{\newblock}{\relax}
\providecommand{\bibinfo}[2]{#2}
\providecommand{\BIBentrySTDinterwordspacing}{\spaceskip=0pt\relax}
\providecommand{\BIBentryALTinterwordstretchfactor}{4}
\providecommand{\BIBentryALTinterwordspacing}{\spaceskip=\fontdimen2\font plus
\BIBentryALTinterwordstretchfactor\fontdimen3\font minus
  \fontdimen4\font\relax}
\providecommand{\BIBforeignlanguage}[2]{{%
\expandafter\ifx\csname l@#1\endcsname\relax
\typeout{** WARNING: IEEEtran.bst: No hyphenation pattern has been}%
\typeout{** loaded for the language `#1'. Using the pattern for}%
\typeout{** the default language instead.}%
\else
\language=\csname l@#1\endcsname
\fi
#2}}
\providecommand{\BIBdecl}{\relax}
\BIBdecl

\bibitem{lecun2015deep}
Y.~LeCun, Y.~Bengio, and G.~Hinton, ``Deep learning,'' \emph{Nature}, vol. 521,
  no. 7553, pp. 436--444, 2015.

\bibitem{zhu_introduction_2009}
X.~Zhu and A.~B. Goldberg, ``Introduction to {Semi}-{Supervised} {Learning},''
  \emph{Synthesis Lectures on Artificial Intelligence and Machine Learning},
  vol.~3, no.~1, pp. 1--130, Jan. 2009.

\bibitem{chapelle2006semi}
O.~Chapelle, B.~Sch{\"o}lkopf, and A.~Zien, ``Semi-supervised learning, ser.
  adaptive computation and machine learning,'' 2006.

\bibitem{cuoco_enhancing_2020}
E.~Cuoco, J.~Powell, M.~Cavaglià, K.~Ackley, M.~Bejger, C.~Chatterjee,
  M.~Coughlin, S.~Coughlin, P.~Easter, R.~Essick, H.~Gabbard, T.~Gebhard,
  S.~Ghosh, L.~Haegel, A.~Iess, D.~Keitel, Z.~Márka, S.~Márka, F.~Morawski,
  T.~Nguyen, R.~Ormiston, M.~Pürrer, M.~Razzano, K.~Staats, G.~Vajente, and
  D.~Williams, ``Enhancing gravitational-wave science with machine learning,''
  \emph{Machine Learning: Science and Technology}, vol.~2, no.~1, p. 011002,
  Dec. 2020.

\bibitem{marianer_semisupervised_2021}
T.~Marianer, D.~Poznanski, and J.~X. Prochaska, ``A semisupervised machine
  learning search for never-seen gravitational-wave sources,'' \emph{Monthly
  Notices of the Royal Astronomical Society}, vol. 500, no.~4, pp. 5408--5419,
  Feb. 2021.

\bibitem{2016PhRvL.116f1102A}
B.~P. {Abbott}, R.~{Abbott}, T.~D. {Abbott} \emph{et~al.}, ``{Observation of
  Gravitational Waves from a Binary Black Hole Merger},'' vol.
  116, no.~6, p. 061102, Feb. 2016.

\bibitem{2015CQGra..32g4001L}
J.~{Aasi}, B.~P. {Abbott}, R.~{Abbott} \emph{et~al.}, ``{Advanced LIGO},''
  \emph{Classical and Quantum Gravity}, vol.~32, no.~7, p. 074001, Apr. 2015.

\bibitem{2015CQGra..32b4001A}
F.~{Acernese}, M.~{Agathos}, K.~{Agatsuma} \emph{et~al.}, ``{Advanced Virgo: a
  second-generation interferometric gravitational wave detector},''
  \emph{Classical and Quantum Gravity}, vol.~32, no.~2, p. 024001, Jan. 2015.

\bibitem{zevin2017gravity}
M.~Zevin, S.~Coughlin, S.~Bahaadini, E.~Besler, N.~Rohani, S.~Allen, M.~Cabero,
  K.~Crowston, A.~Katsaggelos, S.~Larson \emph{et~al.}, ``Gravity spy:
  integrating advanced ligo detector characterization, machine learning, and
  citizen science,'' \emph{Classical and Quantum Gravity}, vol.~34, no.~6, p.
  064003, 2017.

\bibitem{saraicassp}
S.~Bahaadini, N.~Rohani, S.~Coughlin, M.~Zevin, V.~Kalogera, and
  A.~Katsaggelos, ``Joint deep multi-view models for glitch classification,''
  in \emph{The 42nd IEEE International Conference on Acoustics, Speech and
  Signal Processing (ICASSP)}, March 2017.

\bibitem{sarajournal1}
S.~Bahaadini, V.~Noroozi, N.~Rohani, S.~Coughlin, M.~Zevin, J.~Smith,
  V.~Kalogera, and A.~Katsaggelos, ``Machine learning for gravity spy: Glitch
  classification and dataset,'' \emph{Information Sciences}, vol. 444, pp. 172
  -- 186, 2018.

\bibitem{direct}
S.~{Bahaadini}, N.~{Rohani}, A.~K. {Katsaggelos}, V.~{Noroozi}, S.~{Coughlin},
  and M.~{Zevin}, ``Direct: Deep discriminative embedding for clustering of
  ligo data,'' pp. 748--752, Oct 2018.

\bibitem{meng_relational_2017}
Q.~Meng, D.~Catchpoole, D.~Skillicom, and P.~J. Kennedy, ``Relational
  autoencoder for feature extraction,'' in \emph{2017 {International} {Joint}
  {Conference} on {Neural} {Networks} ({IJCNN})}, May 2017, pp. 364--371, iSSN:
  2161-4407.

\bibitem{VAT}
T.~Miyato, S.~Maeda, S.~Ishii, and M.~Koyama, ``Virtual adversarial training: a
  regularization method for supervised and semi-supervised learning,''
  \emph{IEEE transactions on pattern analysis and machine intelligence}, 2018.

\bibitem{goodfellow_generative_2014}
I.~J. Goodfellow, J.~Pouget-Abadie, M.~Mirza, B.~Xu, D.~Warde-Farley, S.~Ozair,
  A.~Courville, and Y.~Bengio, ``Generative {Adversarial} {Networks},''
  \emph{arXiv:1406.2661 [cs, stat]}, Jun. 2014, arXiv: 1406.2661.

\bibitem{qi2016sketch}
Y.~Qi, Y.-Z. Song, H.~Zhang, and J.~Liu, ``Sketch-based image retrieval via
  siamese convolutional neural network,'' in \emph{Image Processing (ICIP),
  2016 IEEE International Conference on}.\hskip 1em plus 0.5em minus
  0.4em\relax IEEE, 2016, pp. 2460--2464.

\bibitem{bahaadini_sara_2018_1476156}
\BIBentryALTinterwordspacing
S.~Bahaadini, V.~Noroozi, N.~Rohani, S.~Coughlin, M.~Zevin, J.~Smith,
  V.~Kalogera, and K.~Aggelos, ``{Machine learning for Gravity Spy: Glitch
  classification and dataset},'' Oct. 2018. [Online]. Available:
  \url{https://doi.org/10.5281/zenodo.1476156}
\BIBentrySTDinterwordspacing

\end{thebibliography}

%








\end{document}